\documentclass[twocolumn,preprint2]{aastex631}

\usepackage{epsfig}
\usepackage{diagbox}
\usepackage{multirow}

\begin{document}

\title{Light curves of the explosion of ONe WD+CO WD merger remnant and type Icn supernovae}

\author[0000-0002-2452-551X]{Chengyuan Wu}
\affiliation{Yunnan Observatories, Chinese Academy of Sciences, Kunming 650216, China}
\affiliation{International Centre of Supernovae, Yunnan Key Laboratory, Kunming, 650216, China}
\email{wuchengyuan@ynao.ac.cn}

\author{Shuai Zha}
\affiliation{Yunnan Observatories, Chinese Academy of Sciences, Kunming 650216, China}
\affiliation{International Centre of Supernovae, Yunnan Key Laboratory, Kunming, 650216, China}

\author{Yongzhi Cai}
\affiliation{Yunnan Observatories, Chinese Academy of Sciences, Kunming 650216, China}
\affiliation{International Centre of Supernovae, Yunnan Key Laboratory, Kunming, 650216, China}

\author[0009-0006-8370-3108]{Zhengyang Zhang}
\affiliation{Yunnan Observatories, Chinese Academy of Sciences, Kunming 650216, China}
\affiliation{International Centre of Supernovae, Yunnan Key Laboratory, Kunming, 650216, China}
\affiliation{University of Chinese Academy of Sciences, Beijing 100049, China}

\author{Yi Yang}
\affiliation{Physics Department and Tsinghua Center for Astrophysics,
Tsinghua University, Beijing, 100084, People’s Republic of China}

\author{Danfeng Xiang}
\affiliation{Physics Department and Tsinghua Center for Astrophysics,
Tsinghua University, Beijing, 100084, People’s Republic of China}

\author{Weili Lin}
\affiliation{Physics Department and Tsinghua Center for Astrophysics,
Tsinghua University, Beijing, 100084, People’s Republic of China}

\author{Xiaofeng Wang}
\affiliation{Physics Department and Tsinghua Center for Astrophysics,
Tsinghua University, Beijing, 100084, People’s Republic of China}

\author{Bo Wang}
\affiliation{Yunnan Observatories, Chinese Academy of Sciences, Kunming 650216, China}
\affiliation{International Centre of Supernovae, Yunnan Key Laboratory, Kunming, 650216, China}

\begin{abstract}

Type Icn supernovae (SNe Icn) are a newly detected rare subtype of interacting stripped-envelope supernovae which show narrow P-Cygni lines of highly ionized carbon, oxygen, and neon in their early spectra due to the interactions of the SNe ejecta with dense hydrogen- and helium-deficient circumstellar material (CSM). It has been suggested that SNe Icn may have multiple progenitor channels, such as the explosion of carbon-rich Wolf–Rayet stars, or the explosion of stripped-envelope SNe which undergo binary interactions. Among the SNe Icn, SN 2019jc shows unique properties, and previous work inferred that it may stem from the ultra-stripped supernova, but other possibilities still exist. In this work, we aim to simulate the light curves from the explosions of oxygen-neon and carbon-oxygen double white dwarf (WD) merger remnants, and to further investigate whether the corresponding explosions can appear as some particular SNe Icn. We generate the light curves from the explosive remnants and analyse the influence of different parameters on the light curves, such as the ejecta mass, explosion energy, mass of $^{\rm 56}{\rm Ni}$ and CSM properties. Comparing our results with some SNe Icn, we found that the light curves from the explosions of double WD merger remnants can explain the observable properties of SN 2019jc, which inferred that this special SN Icn may have a different progenitor. Our results indicated that double WD merger may be an alternative model in producing at least one of the SNe Icn.

\end{abstract}


\section{Introduction} \label{sec:intro}

It is widely believed that massive stars with initial masses greater than $8-12{M}_{\odot}$ will end their lives exploding as core collapse supernovae (CCSNe), which can be classified into different types from their spectral properties (e.g., \citealt{1997ARA&A..35..309F}; \citealt{2017hsn..book..195G}). CCSNe that show hydrogen (H) lines in their spectra are classified as Type II SNe, while for those that show complete absence of H features in their spectra, or show evidence of H only at early times are called H-poor or stripped-envelope SNe (SESNe), including types Ib, Ic, and IIb SNe. Some CCSNe show evidence for interactions between the SNe ejecta and circumstellar material (CSM) are named as interacting SNe. The interacting SNe that exhibit bright and narrow Balmer lines of H in their spectra are classified as Type IIn SNe, while the events display narrow lines of helium (He) in their spectra without H, indicating the presence of a H-poor but He-rich CSM around the exploding stars, are named as Type Ibn SNe. Recently, an even more extreme subtype of interacting SESNe, named as Type Icn supernovae (SN Icn) has been discovered (e.g., \citealt{2021arXiv210807278F}; \citealt{2022Natur.601..201G}). The early-time spectra of SNe Icn are blue with narrow P-Cygni lines of highly ionized carbon (C), oxygen (O) and neon (Ne), instead of He. These SNe show rapid photometric evolution with a high peak brightness. After the maximum light, as the SN ejecta becomes optically thin, their spectra appear more similar to normal SNe Ibc or SNe Ibn but still remain different (e.g. \citealt{2022Natur.601..201G}; \citealt{2022ApJ...927..180P}; \citealt{2023MNRAS.523.2530D}). 

Up to now, there are only five SNe Icn have been reported, i.e., SN 2019hgp, SN 2019jc, SN 2021csp, SN 2021ckj and SN 2022ann (e.g., \citealt{2021arXiv210807278F}; \citealt{2022ApJ...927..180P}; \citealt{2022Natur.601..201G}; \citealt{2022ApJ...938...73P}; \citealt{2023A&A...673A..27N}; \citealt{2023MNRAS.523.2530D}). The origins of SNe Icn are still under debate. \cite{2022Natur.601..201G} proposed a Wolf-Rayet (WR) star to be the progenitor of SN 2019hgp based on its observational properties, and suggested a possibility that the differences between Type Ibn and Type Icn SNe arise from their different types of WR star progenitors, i.e., He- and nitrogen (N)-rich WN stars for Type Ibn SNe and C-rich WC stars for Type Icn SNe. \cite{2022ApJ...927..180P} suggested that a WR star directly collapse to a black hole can launch a subrelativistic jet, and the jet interacts with the dense CSM may explain the observational properties of SN 2021csp. \cite{2023A&A...673A..27N} suggested that SN 2021csp and SN 2021ckj may have aspherical ejecta and spherical CSM, and the different observational properties between these two SNe arise from the viewing angle effects (see also \citealt{2021arXiv210807278F}), i.e., the SN 2021ckj is observed from a direction close to the jet pole, while SN 2021csp is observed from an off-axis direction. \cite{2023MNRAS.523.2530D} found that the CSM of SN 2022ann only has a velocity of about $800\,{\rm {km}}/{\rm s}$, which is inconsistent with the typical velocity of WR star winds. Moreover, the host galaxy of SN 2022ann is a low-mass dwarf galaxy with low star formation rate and low averaged metallicity, suggesting that the progenitor of SN 2022ann has low metallicity. This implies that it is difficult to fully stripped the H/He envelope from its progenitor through stellar wind (massive WR star may not be the progenitor of SN 2022ann). Instead, they suggested a binary-stripped progenitor for SN 2022ann, rather than a single massive WR progenitor. \cite{2022ApJ...938...73P} analyzed a sample of four SNe Icn, including SN 2019hgp, SN 2019jc, SN 2021ckj, and SN 2021csp. They found that SN 2019jc shows different features comparing with other SNe Icn in the following aspects: first, this SN is unique in having a light curve that is both faster evolving and fainter than those of other SNe Icn; secondly, SN 2019jc is located at the outer edge of its host galaxy, where the star formation rate density is low. Based on a circumstellar interaction model, \cite{2022ApJ...938...73P} fitted the bolometric light curves of these four SNe Icn, and found that SN 2019jc has low explosion energy and low ejecta mass, suggesting a low mass, ultra-stripped SN as its progenitor. However, for other SNe Icn, the WR star progenitors or binary systems may better explain their observational properties.

Recently, \cite{2023ApJ...944L..54W} investigated the evolution of the post-merger remnants from the coalescence of ONe and CO WD pairs. They found that the merger remnants could evolve to H/He-deficient giant-like stars due to the shell carbon burning. The merger remnants with masses greater than $1.95{M}_{\odot}$ can explode as electron-capture SNe (ECSNe), and could produce CO-rich CSM resulted from the stellar wind during their giant phases. The interactions between the SNe ejecta and CSM may produce narrow C, O and Ne emission lines similar to SNe Icn, making the double WD merger is able to be a potential channel in producing SNe Icn. Even though some of the SNe Icn observed at early times show He emission lines (for example, the spectrum of SN 2019jc at maximum light shows He II $\lambda4686$ feature), suggesting that the CSM surrounding SNe Icn may not fully devoid of He. Note that there could exist a He surface layer on the CO WD, which causes the merger remnant involving at least one CO WD could contain amount of He (e.g., \citealt{2014MNRAS.438...14D}; \citealt{2021ApJ...906...53S}; \citealt{2023ApJ...944L..54W}). The subsequent evolution of such merger remnant may produce CSM that is dominated by C, O and Ne but not completely devoid of He through stellar wind, which means that it is possible to detect He emission lines from the early-time spectra of some SNe Icn, if these SNe are stemmed from the double WD merger channel. Since the optical light curves from the explosions of the massive double WD merger remnants have not been well studied, in this work, we are going to simulate the light curves from the explosions of merger remnants, and then to compare them with some SNe Icn.

The rest of the article is organized as follows: in Sect.\,2, we introduce the method in generating the light curves of exploding double WD merger remnants. In Sect.\,3, we analyse how different sets of parameters in influencing the light curves. In Sect.\,4, we compare our simulated light curves to some SNe Icn, and discuss whether our model can explain SNe Icn. Finally, the summary are provided in Sect.\,5.

\section{Model setups} \label{sec:method}

As investigated by previous works, ONe+CO WD merger remnants with masses greater than $1.95{M}_{\odot}$ will end their lives as ECSNe (e.g., \citealt{2023ApJ...944L..54W}). The collapse of the ONe core releases a large amount of gravitational potential energy, which can be transformed to thermal and kinetic energy to power a bright transient. In order to calculate the light curves from the explosions of double WD merger remnants, we take the stellar structures at their final evolutionary stages (just prior to the core collapse) from \cite{2023ApJ...944L..54W} as our initial models. Then we evolve the models from a few seconds after the central explosions triggered by core collapse to a time just before the outgoing shocks reach the stellar surface by employing the Modules for Experiments in Stellar Astrophysics ($\tt{MESA}$, version 12778; e.g., \citealt{2011ApJS..192....3P}; \citealt{2013ApJS..208....4P}; \citealt{2015ApJS..220...15P}; \citealt{2018ApJS..234...34P}; \citealt{2019ApJS..243...10P}). Subsequently, we use radiation hydrodynamics code $\tt{STELLA}$ (e.g., \citealt{1993A&A...273..106B}; \citealt{1996AstL...22...39B}; \citealt{1998ApJ...496..454B}; \citealt{2000ApJ...532.1132B}; \citealt{2006A&A...453..229B}; \citealt{2011AstL...37..194B}) to evolve the models from the shock breakout to several tens of days to TELLAgenerate the light curves.

In detail, our approach is as follows. First, we consider the most massive merger remnant described in \cite{2023ApJ...944L..54W} as our initial model (i.e., $2.25{M}_{\odot}$, resulted from an $1.20{M}_{\odot}$ ONe WD merge with an $1.05{M}_{\odot}$ CO WD). The central core temperature of this model achieved ${\rm {log}}({T}_{\rm c}/{\rm K})=9.3$ when core collapse occurs, and the radius of the remnant is about $377{R}_{\odot}$ at that moment. To investigate the effect of ejecta mass on the light curves, we remove different central sections from the initial model. We assume three mass cuts, i.e., $1.42{M}_{\odot}$ (just above the CO burning layer), $1.65{M}_{\odot}$ and $1.89{M}_{\odot}$, corresponding to the ejecta masses equal to $0.83{M}_{\odot}$, $0.60{M}_{\odot}$ and $0.36{M}_{\odot}$, respectively. The mass profile and density structure of the initial model as well as the positions of different mass cuts are shown in panel (a) and (b) of Fig.\,1.

Secondly, we put the structures above the mass cuts into the explosion
calculations. According to the numerical explosion simulations, the explosion energy of ECSNe is in an order of magnitude of ${10}^{50}\,{\rm {erg}}$ (e.g., \citealt{2006A&A...450..345K}; \citealt{2016MNRAS.461.2155M}; \citealt{2022MNRAS.513.1317Z}). To explore the influence of explosion energy on the light curves, we inject different energies (i.e., $0.05\,{\rm B}$, $0.10\,{\rm B}$ and $0.20\,{\rm B}$; $1\,{\rm B}\equiv{10}^{51}\,{\rm {erg}}$) into a thin layer of about $0.01{M}_{\odot}$ at the inner boundary of the remaining envelope within $5$ milliseconds.

Thirdly, we calculate the shock propagating process from the innermost shell of the envelope towards the surface. Since it is important to have enough unshocked material for the $\tt{STELLA}$ code to calculate the light curves, we stop our $\tt{MESA}$ simulations once there is $\sim0.05{M}_{\odot}$ unshocked material remaining between the shock and the surface. During the shock propagating process, we ignore all the nuclear reactions to keep the elemental abundance distributions in the envelope remain unchanged. Note that $^{\rm 56}{\rm Ni}$ can be generated by the intense nuclear burning behind the shock near the core-envelope boundary just after core collapse, and numerical simulations indicated that the mass of $^{\rm 56}{\rm Ni}$ produced in ECSNe and accretion-induced-collapse events would be in the order of magnitude of ${10}^{-3}$ to ${10}^{-4}\,{M}_{\odot}$ (e.g., \citealt{2006ApJ...644.1063D}; \citealt{2016MNRAS.461.2155M}). To explore how the mass of $^{\rm 56}{\rm Ni}$ effects on the light curves, we artificially add different masses of $^{\rm 56}{\rm Ni}$ (i.e., $0.0001{M}_{\odot}$, $0.001{M}_{\odot}$ and $0.01{M}_{\odot}$) in the envelope. This goal is achieved by reducing the mass of $^{\rm 16}{\rm O}$ and replacing with $^{\rm 56}{\rm Ni}$. Besides, previous works suggested that the $^{\rm 56}{\rm Ni}$ mixing process may influence the sharp of light curves (e.g., \citealt{2020MNRAS.497.1619M}). In our simulations, we considered two different degrees of mixing. In the ``full mixing'' model, $^{\rm 56}{\rm Ni}$ is uniformly mixed in the entire ejecta, while in the ``half mixing'' model, the outer edge of the layer where $^{\rm 56}{\rm Ni}$ is mixed is in the center of the envelope. The elemental abundance distribution profile of the ejecta with $0.001{M}_{\odot}$ of $^{\rm 56}{\rm Ni}$ is shown in panel (c) of Fig.\,1.

Finally, after the outer edge of the shock reached the desired position, we save the $\tt{STELLA}$ data by adopting ``stella$\_$num$\_$points=$400$'', where $400$ is a recommended value in $\tt{MESA}$ default. Note that the stellar winds of the merger remnants prior to their explosions may form CSM surround the progenitors, and the ejecta interacting with CSM could transform kinetic energy into thermal energy and thus influence the light curves. To investigate this effect, we attach the CSM on the top of the progenitors, and use ``stella$\_$nz$\_$extra=$40$'' to save the data of CSM properties. We consider the wind-like CSM. Based on the evolutionary results from the previous work (e.g. \citealt{2023ApJ...944L..54W}), the merger remnants could have enormous wind mass-loss rates just prior to the SN explosions (the time intervals between the significant wind mass-loss processes and the SN explosions are within tens of days to several years, which are influenced by the total masses of the merger remnants and the wind mass-loss prescriptions), which means that the outer radii of the CSM surrounding the SNe are not too large. For the wind-like CSM, the wind density parameter ($w$) is very useful in directly estimating the density of the CSM (${\rho}_{\rm {CSM}}$), i.e.,
\begin{equation}
    {w}=\frac{\dot{M}}{{v}_{\rm {wind}}},
\end{equation}
and 
\begin{equation}
    {\rho}_{\rm {CSM}}({\rm r})=\frac{w}{4\pi{r}^{-2}},
\end{equation}
where, $\dot{M}$ is the mass-loss rate of the progenitors, ${v}_{\rm {wind}}$ is the velocity of the stellar wind, and ${r}$ is the radius of CSM (e.g., \citealt{2017hsn..book..843B}). Since the wind velocities and mass-loss rates of such H/He-deficient giants are not well constrained from the observations, in the current work, we consider various of wind velocities (${v}={10}\,{\rm km}/{\rm s}$, ${100}\,{\rm km}/{\rm s}$ and ${1000}\,{\rm km}/{\rm s}$, corresponding to the typical wind velocities of RGB/AGB stars, yellow supergiants and WR stars, respectively) (e.g., \citealt{2017hsn..book..403S}), wind density parameters (${3.15}\times{10}^{17}-{3.15}\times{10}^{19}\,{\rm g}/{\rm {cm}}$, depending on the evolution of the progenitor stars under different wind mass-loss parameters) and total masses of CSM (${M}_{\rm {CSM}}={0.05}$ and $0.25{M}_{\odot}$). The density and velocity profiles of the remnants and CSM (an example that we assume the enormous wind mass-loss rate, i.e., $0.5{M}_{\odot}/{\rm {yr}}$, occurs $0.1$ year prior to the SN explosion) are provided in panel (d) of Fig.\,1.

\begin{figure*}
\begin{center}
\epsfig{file=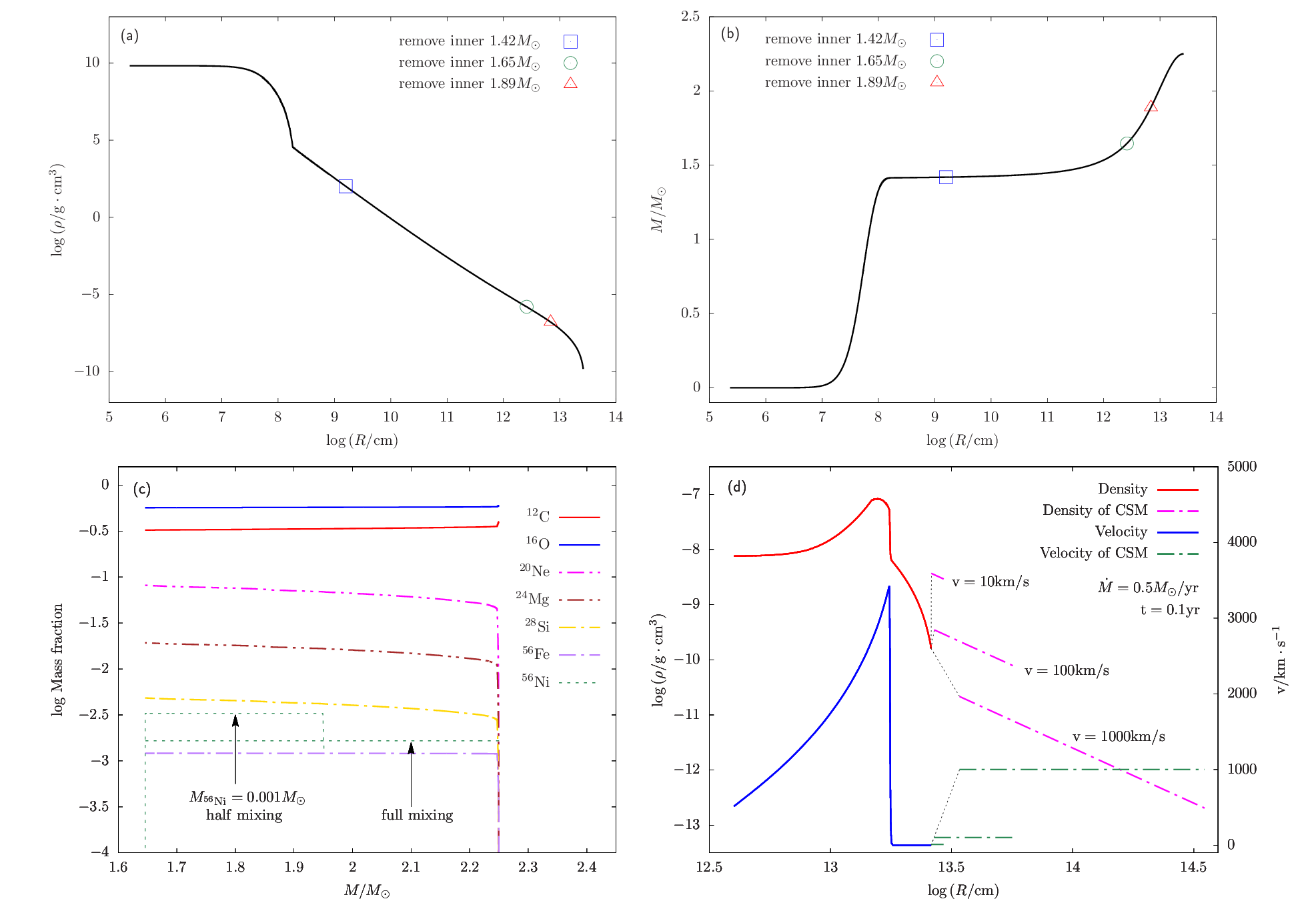,angle=0,width=16.2cm}
 \caption{Properties of the merger remnant before light curves calculations. Panel (a) and (b) represent the density profile and the enclosed mass as the function of radius of the merger remnant before explosion. Blue squares, green circles and red triangles represent the different mass cuts before energy injection. Panel (c) and (d) represent the elemental abundance profile and the density and velocity profiles of the $0.6{M}_{\odot}$ ejecta at the moment when shock wave reached the position at $0.05{M}_{\odot}$ beneath the surface. Panel (c) also shows the elemental abundance distribution in considering $^{\rm 56}{\rm Ni}$ with half and full mixing. Panel (d) also shows the properties of CSM with different wind velocities (wind mass-loss rate of $0.5{M}_{\odot}/{\rm {yr}}$ lasts for $0.1$ year prior to the SN explosion).}
  \end{center}
    \label{fig: 1}
\end{figure*}

\section{Bolometric Light Curves} \label{LC}

We use the multi-group radiation hydrodynamics code $\tt{STELLA}$ to follow the evolution of the explosion models described above through shock breakout to $70$ days after explosion. $\tt{STELLA}$ is able to simulate SN evolution at early times, and can also solve the shock breakout and interaction between ejecta and CSM. In the calculation, SN explosion is regarded as a thermal bomb locating at the inner boundary of the ejecta. $\tt{STELLA}$ solves the radiative transfer equations in the intensity momentum approximation in each frequency bin. We use $40$ frequency bins, which is enough to generate bolometric light curves. The opacity is computed based on over $153000$ spectral lines from \cite{1995all..book.....K} and \cite{1996ADNDT..64....1V}. Some simulated outputs from our $\tt{STELLA}$ simulations are publicly available on Zenodo at doi:10.5281/zenodo.11066927.

We present the light curves from different progenitors described above in Fig.\,2, while the evolution of blackbody temperature, density, radius, velocity and opacity of a specific model (i.e., ${M}_{\rm {ej}}=0.60{M}_{\odot}$, ${E}=0.1\,{\rm B}$, without $^{\rm 56}{\rm {Ni}}$, without CSM) are presented in Fig.\,3. The overall evolutionary characteristics of the luminosity are as follows (four evolutionary phases). (1) Rapid increasing phase: at the beginning of the evolution, the ejecta is heated and ionized by the outwardly propagating explosion shock wave. (2) Rapid declining phase: after the shock breakout, the evolution of the bolometric luminosity is powered by the diffusion of thermal energy depositing in the ejecta, and during this stage, the luminosity declines quickly. (3) Plateau phase: as the ejecta expands and cools, the temperature of the material gradually drops below the recombination temperature and become largely transparent. The composition of the ejecta in our model is mixed mainly by $^{\rm 12}{\rm C}$, $^{\rm 16}{\rm O}$, $^{\rm 20}{\rm {Ne}}$ and $^{\rm 24}{\rm {Mg}}$. For this mixture, the recombination temperature is about $6000\,{\rm K}$. For the higher temperature (${T}>6000\,{\rm K}$), the opacity of the mixture is greater than about ${0.04}\,{\rm {cm}}^{2}\,{\rm {g}}^{-1}$. When the temperature drops below $6000\,{\rm K}$, oxygen recombines to neutral and the opacity drops dramatically, which is similar to the O-Ne-Mg mixture (e.g., \citealt{2014MNRAS.438..318K}). As the outer layers of the ejecta begin to recombines, a sharp ionization front develops in the ejecta. At this moment, the photosphere stays at the recombination front and the luminosity decline rate becomes significantly smaller than before, resulting in the appearance of plateau-like feature. (4) Late phase: the plateau phase lasts for about $10$-$25$ days depending on the initial structures of the progenitors, and after which the recombination wave reaches the center of the ejecta, exhausting the energy stored in the ejecta and the light curve drops off extremely rapid, making the end of the plateau phase.

The duration of the plateau phase is dominated by the dissipation rate of the energy stored in the ejecta. From Fig.\,2, we can see that the ejecta mass and the explosion energy can significantly effect the plateau luminosity and the duration of plateau phase (panel a and b). Under the same explosion energy, a more massive ejecta mass leads to a lower plateau luminosity. This is because when the recombination front developed in the envelope, the ejecta usually have similar photosphere temperature, but a more massive ejecta has lower expansion velocity, which causes a smaller photosphere radius. Besides, a lower expansion velocity makes the energy deposited in the envelope needs more time to be exhausted, hence, the plateau phase is longer for a more massive ejecta mass. The effect of explosion energy on the light curves is similar to that of ejecta mass, i.e., a higher explosion energy leads to a more luminous but shorter plateau phase, and vice versa.

After the SN explosion, the nuclear burning behind the shock wave is able to produce iron-group elements such as $^{\rm 56}{\rm Ni}$ in the ejecta. The radioactive decay of $^{\rm 56}{\rm Ni}$ could heat the envelope and thus increase the bolometric luminosity. In panel (c) of Fig.\,2, we compare the light curves of the progenitor models with different masses of $^{\rm 56}{\rm Ni}$. Various masses of $^{\rm 56}{\rm Ni}$ cannot influence the light curves during the plateau phase significantly. About $15$ days after explosion, the radioactive decay of isotopes gradually dominates the evolution of the luminosity, and a ten times of $^{\rm 56}{\rm Ni}$ mass will cause an order of magnitude increasing in the tail bolometric luminosity. Theoretically, if $^{\rm 56}{\rm Ni}$ is entirely mixed in the ejecta, the heating caused by radioactive decay in the outer layers of the ejecta becomes efficient, which leads to the increase in luminosity occurs earlier than that if $^{\rm 56}{\rm Ni}$ is only located at the inner portion of the ejecta. However, due to the low $^{\rm 56}{\rm Ni}$ mass generated from ECSNe, the plateau luminosity is still dominated by the energy deposited in the envelope, which means that the degree of $^{\rm 56}{\rm Ni}$ mixing may not effect the light curves significantly in the current situation.

After the merger of double WDs, the remnant can evolve to giant phase, and during which the H/He-deficient CSM is formed due to the stellar wind. The interaction between ejecta and CSM can transform kinetic energy into thermal energy and thus alter the features of the light curves. The interacting process is mainly influenced by the properties of the CSM such as its radius and its density distribution. In panel (d) of Fig.\,2, we present the light curves of the explosion remnants with different CSM properties. Since the wind mass-loss rates and wind velocities of such merger remnants are not well constrained, we artificially added $0.05{M}_{\odot}$ or $0.25{M}_{\odot}$ of CSM surrounding the progenitor. The density profile of the CSM is dominated by the wind density parameters, and in these simulations, the parameter $w$ is in the range between $3.15\times{10}^{17}$ and $3.15\times{10}^{19}\,{\rm g}/{\rm {cm}}$, which is related to the wind mass-loss histories of the progenitors. We consider three different wind velocities, i.e., ${v}=10\,{\rm km}/{\rm s}$, ${v}=100\,{\rm km}/{\rm s}$ and ${v}=1000\,{\rm km}/{\rm s}$, corresponding to the typical wind velocity of RGB/AGB stars, yellow supergiants and WR stars, respectively (e.g., \citealt{2017hsn..book..403S}). From the panel (d) of Fig.\,2, we can see that the early bolometric luminosity becomes higher when the wind density parameter of CSM is higher because of the extra radiation from CSM interaction. A longer CSM interaction resulted from the larger radius of the CSM causes the delayed appearance of peak luminosity in the light curves. Afterwards, all the light curves restore to the similar evolutionary paths, indicating the end of the CSM interaction.

\begin{figure*}
\begin{center}
\epsfig{file=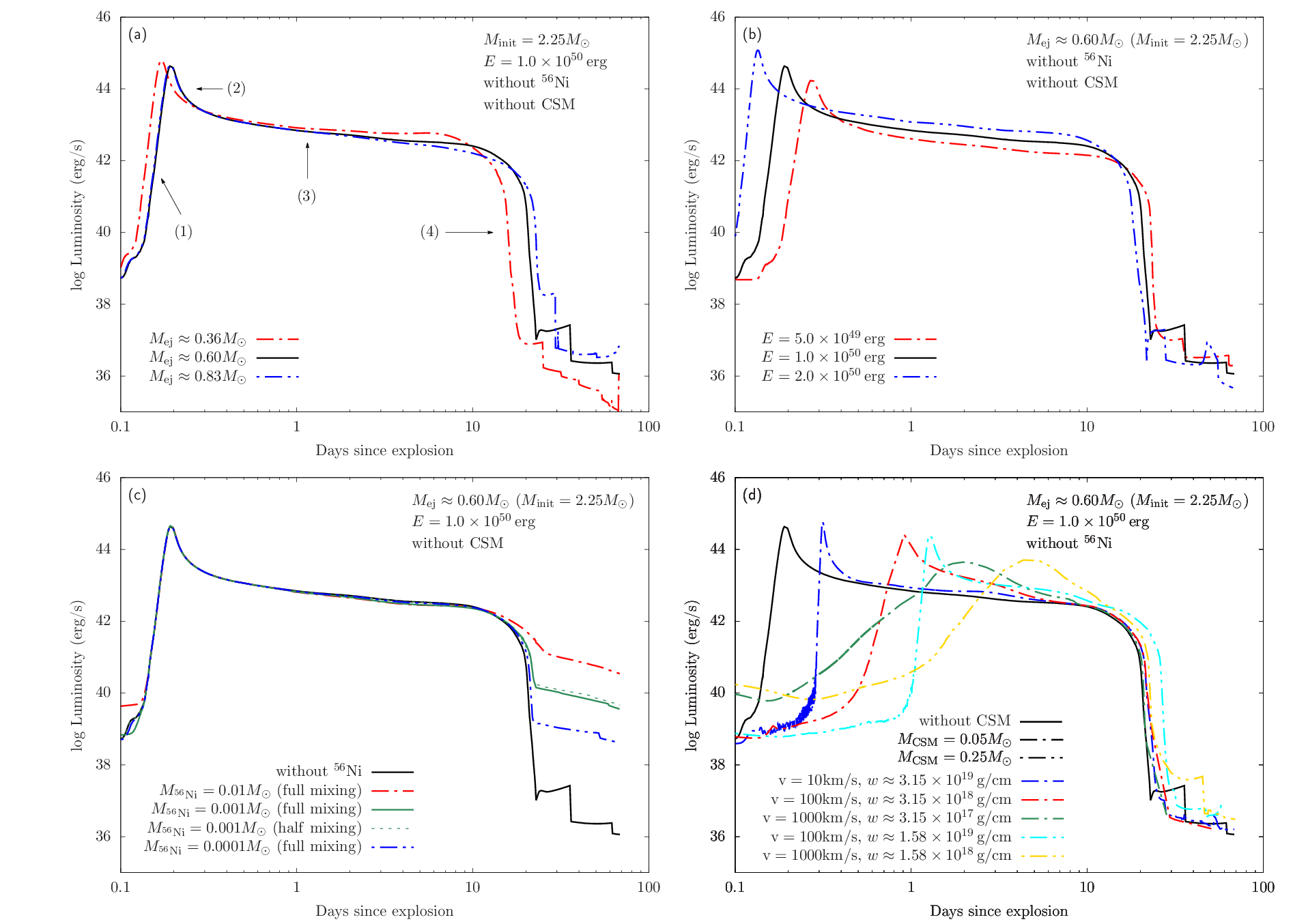,angle=0,width=16.2cm}
 \caption{Light curves from the explosions of double WD merger remnants with different initial parameters, demonstrating how the parameters alter the light curves. Panel (a): the effect of ejecta mass (${M}_{\rm ej}$). Panel (b): the effect of explosion energy (${E}$). Panel (c): the effect of $^{\rm 56}{\rm Ni}$ mass (${M}_{^{\rm 56}{\rm Ni}}$) and mixing. Panel (d): the effect of CSM properties. Number (1)-(4) in panel (a) represent the four evolutionary phases mentioned in Sect.\,3.}
  \end{center}
    \label{fig: 2}
\end{figure*}

\begin{figure*}
\begin{center}
\epsfig{file=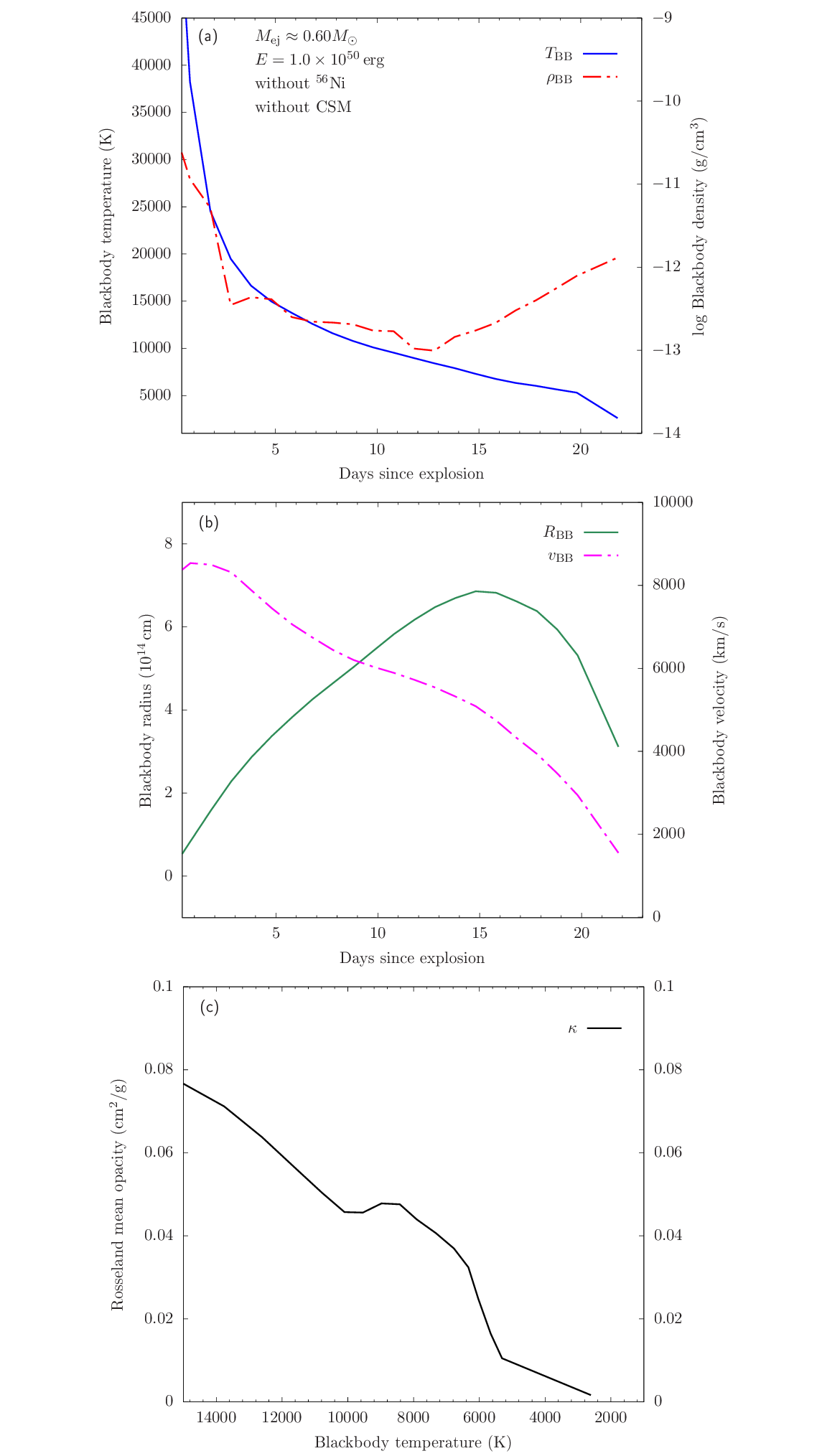,angle=0,width=11.8cm}
 \caption{Additional properties of the merger remnant (${M}_{\rm ej}=0.60{M}_{\odot}$, ${E}=1.0\times{10}^{50}\,{\rm erg}$) after explosion. Panel (a) and (b): blackbody temperature, density, radius and velocity as the function of time. Panel (c): Rosseland mean opacity at the photosphere as a function of temperature.}
  \end{center}
    \label{fig: 3}
\end{figure*}

\section{Comparing to observation} \label{Observation}

To date, five SNe Icn have been reported. According to the analyses from the previous works, it seems that this new type of SNe may have multiple formation channels (e.g., \citealt{2022Natur.601..201G}; \citealt{2022ApJ...927..180P}; \citealt{2022ApJ...938...73P}). In order to investigate whether any SNe Icn may have originated from the post-merger remnants resulted from the coalescence of ONe and CO WD pairs, we obtain some observational data of three SNe Icn, i.e., SN 2019hgp, SN 2019jc and SN 2021csp from \cite{2022ApJ...938...73P}, and compare our models to the observational data. The properties of the progenitors and the CSM are based on the evolutionary results of $2.25{M}_{\odot}$ double WD merger remnant provided in \cite{2023ApJ...944L..54W}. During the final evolutionary stages of the merger remnants, there exist strong wind mass-loss processes, which lead to the decrease in the radii of the progenitors and the formation of dense CSM. However, note that it is difficult to accurately estimate the progenitor radius and the mass of the CSM because of the uncertainty in wind mass-loss rate. Hence, in our models, we assume the enormous stellar wind decreases the radius of the progenitor down to $100{R}_{\odot}$, making the ejecta mass equals to $0.468{M}_{\odot}$, and the dense CSM mass equals to $0.025{M}_{\odot}$. Besides, we assume that the wind velocity equals to $1000\,{\rm km}/{\rm s}$, which is consistent with the observed CSM velocity of SNe Icn. We consider different explosion energies (from $E=1.0\times{10}^{50}$ to $4.0\times{10}^{50}\,{\rm {erg}}$) and $0.01{M}_{\odot}$ of $^{\rm 56}{\rm Ni}$ in the ejecta to simulate the corresponding light curves. For comparison, we also simulate the light curves with the same progenitor/CSM properties (${R}=100{R}_{\odot}$, ${M}_{\rm {ej}}=0.468{M}_{\odot}$ and ${E}=1.0\times{10}^{50}\,{\rm {erg}}$) but without $^{\rm 56}{\rm Ni}$ in the ejecta or even without the interaction between ejecta and CSM.

\begin{figure*}
\begin{center}
\epsfig{file=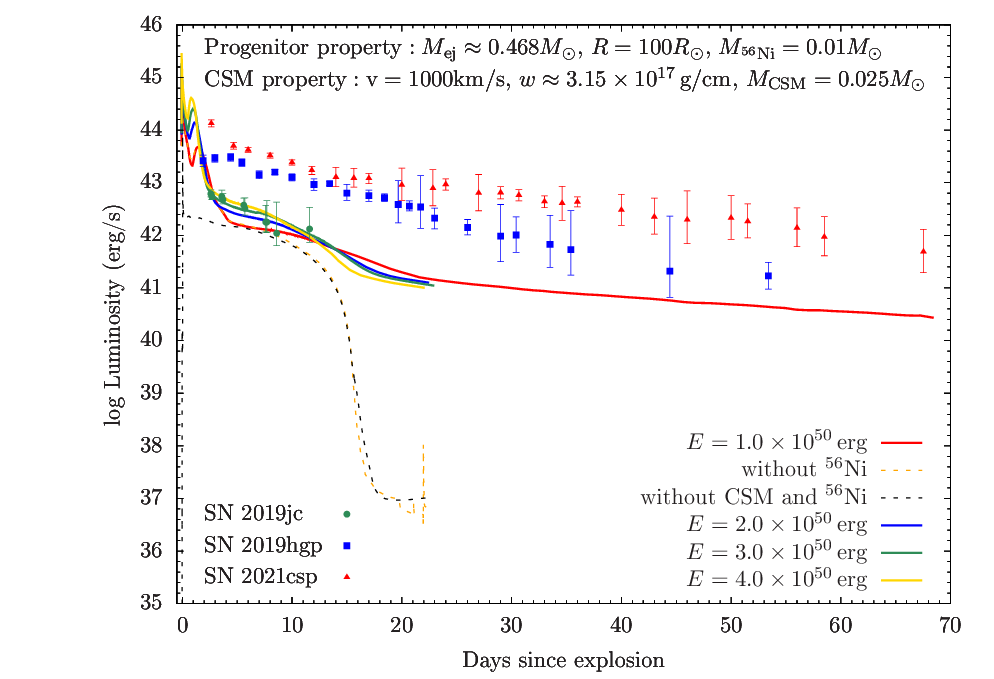,angle=0,width=14.2cm}
 \caption{The comparison between light curves from the explosions of double WD merger remnants and the bolometric light curves of three SNe Icn. In the light curve simulations, the radius of progenitor is ${R}=100{R}_{\odot}$, the ejecta mass is ${M}_{\rm {ej}}\approx0.468{M}_{\odot}$, and the mass of nickel ${M}_{^{\rm 56}{\rm {Ni}}}=0.01{M}_{\odot}$ fully mixed into the ejecta. For the property of CSM, the wind velocity is ${v}=1000{\rm {km}}/{\rm s}$, wind density parameter is ${w}\approx3.15\times{10}^{17}\,{\rm g}/{\rm {cm}}$, and the mass of CSM is ${M}_{\rm {CSM}}=0.025{M}_{\odot}$. Solid lines with different colors represent the light curves from our simulations with different explosive energies (the CSM interaction and radioactive decay from $^{\rm 56}{\rm {Ni}}$ are considered). Dotted lines represent the simulated light curves with explosive energy ${E}=0.1\,{\rm B}$, but without considering radioactive decay from $^{\rm 56}{\rm {Ni}}$ (orange dotted line) and without considering both CSM interaction and radioactive decay from $^{\rm 56}{\rm {Ni}}$ (black dotted line). For some of the models, the simulations terminate at $23$ days after explosions in order to save the calculation time. Green circles, blue squares and red triangles with error bars represent the bolometric light curves of SN 2019jc, SN 2019hgp and SN 2021csp, respectively.}
  \end{center}
    \label{fig: 4}
\end{figure*}

The comparison of our simulated light curves with these three SNe is presented in Fig.\,4, while the comparison of the evolution of blackbody temperature and radius are presented in panel (a) and (b) of Fig.\,5, and the multi-band light curves from two of our models (${E}=2.0\times{10}^{50}$ and $4.0\times{10}^{50}\,{\rm {erg}}$) are provided in panel (c) and (d) of Fig.\,5. The bolometric luminosities and blackbody properties of each objects are estimated through the extensive multi-band observations (for more detail, see \citealt{2022ApJ...938...73P}). From Fig.\,4, we can see that the bolometric luminosities from the explosion of double WD merger remnants are significantly lower that that of SN 2019hgp and SN 2021csp. This implies that these two SNe should have more massive ejecta masses and higher explosion energies, indicating that they may related to the explosions of massive WR stars. Interestingly, the light curves from our models with explosion energy ${E}=2.0\times{10}^{50}-4.0\times{10}^{50}\,{\rm {erg}}$ provide reasonable fits to the bolometric luminosity of SN 2019jc\footnote{Note that the progenitor/CSM properties in our models are different with the estimated values from \cite{2022ApJ...938...73P}. However, it is difficult to constrain some properties such as ejecta mass, CSM mass, progenitor radius from the observation directly, but the CSM velocity can be inferred from the spectra. Thus, in our models, we assume the CSM velocity is $1000\,{\rm km}/{\rm s}$, which is consistent with SNe Icn, but other parameters are related to the evolution of the progenitors.}, which means that this SN may have a different progenitor from other SNe Icn. Furthermore, SN 2019jc exploded on the outskirts of the nearby spiral galaxy, where the star formation rate density is low. The low star formation rate density argues against the progenitor with high zero-age main-sequence mass. The progenitors of WDs are low mass main-sequence stars and the SNe from the double WD mergers could form in any positions of their host galaxies, which implies that the host galaxy environment of SN 2019jc is also consistent with double WD merger channel. However, it is still too early to make a conclusion that SN 2019jc is related to double WD merger. On one hand, although the luminosities of B, V, R, I bands are consistent with SN 2019jc, the evolution of multi-band light curves obtained from our models are slower than that of SN 2019jc, which is probably because of the discrepancy in the photosphere opacity resulted from the difference of ejecta compositions or the radius of the progenitor. On the other hand, SN 2019jc was not observed until approximately 3 days after explosion and disappeared approximately 12 days after explosion, which cannot constrain its early-time and subsequent evolution. For the early-time evolution, our results imply that ultraviolet (UV) or X-ray radiation may have major contribution to the bolometric light curves. For the late-time evolution, the existence of $^{\rm 56}{\rm Ni}$ in the ejecta can significantly influence the luminosity after the plateau phase. Hence, the early-time UV or X-ray observations and the late-time observational data are needed in constraining the progenitor channel of SN 2019jc. Currently, even though we cannot provide any better limitations on the origin of SN 2019jc depending on the limited data, we suggest that the explosion of double WD merger remnant may be a new model in explaining the observational properties of SN 2019jc, but other progenitor channels such as ultra-stripped SN or stripped envelope of ECSN (e.g., \citealt{2013ApJ...778L..23T}; \citealt{2015MNRAS.451.2123T}; \citealt{2016MNRAS.461.2155M}) are still possible.

\begin{figure*}
\begin{center}
\epsfig{file=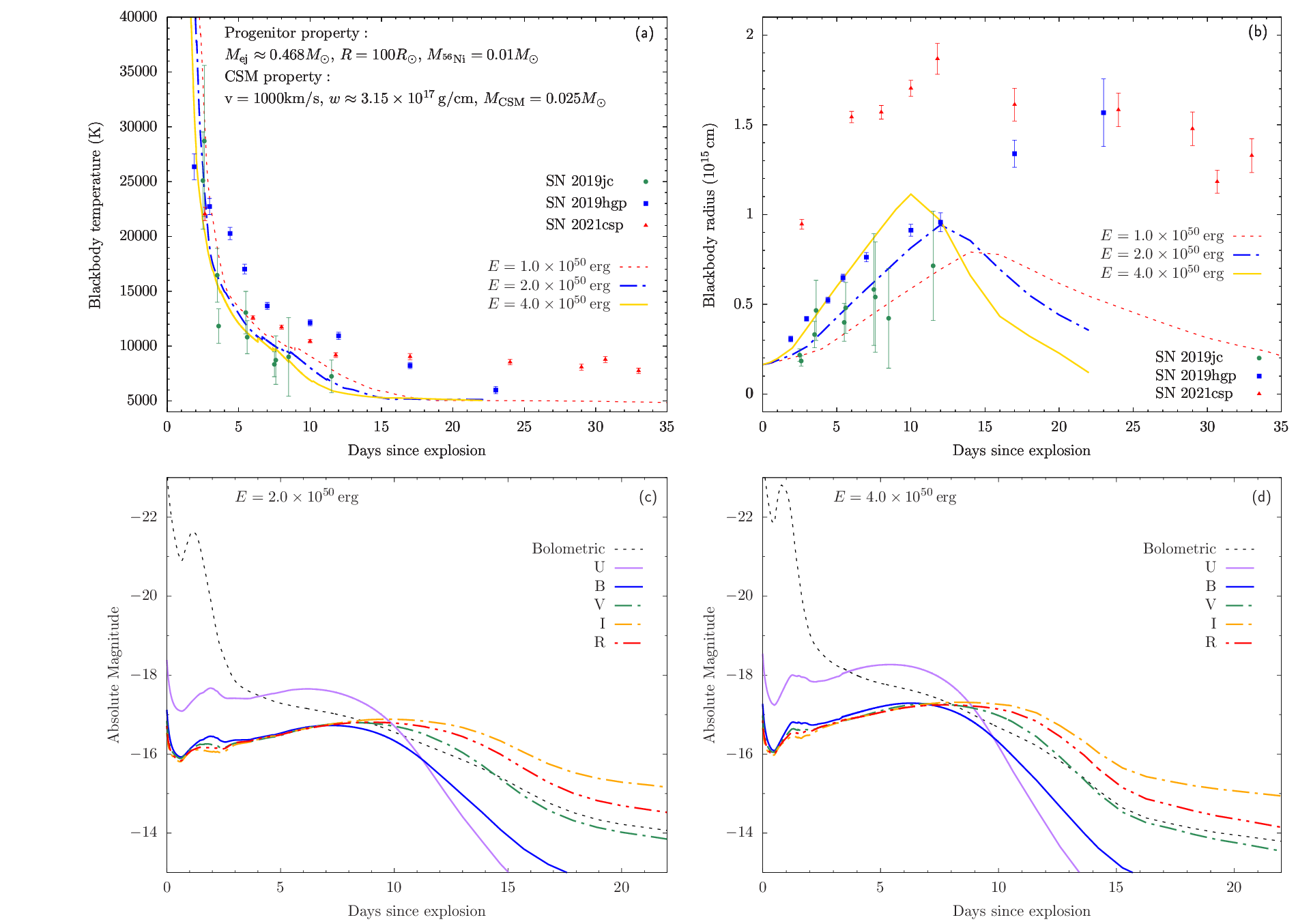,angle=0,width=17.2cm}
 \caption{Panel (a) and (b): similar to Fig.\,4, but for the evolution of blackbody temperature (panel a) and blackbody radius (panel b). Panel (c) and (d): multi-band light curves for double WD merger remnants with different explosion energies.}
  \end{center}
    \label{fig: 5}
\end{figure*}

\section{Summary} \label{Summary}

In this work, we simulated the light curves from the explosion of ONe+CO WD merger remnants. We found that the light curves decline quickly after the shock breakout and enter into the plateau phases resulted from the recombination of ionized carbon and oxygen. After the energy stored in the envelope are exhausted, the light curves drop off rapidly. In the simulations, we analysed the influences of different parameters such as ejecta mass, explosion energy, $^{\rm 56}{\rm Ni}$ mass and different properties of CSM. The ejecta mass and explosion energy can effect the plateau bolometric luminosities and their durations. A higher explosion energy or a lower ejecta mass leads to a more luminous but shorter plateau phase, and vice versa. The $^{\rm 56}{\rm Ni}$ mass can effect the tail bolometric luminosity. A ten times of $^{\rm 56}{\rm Ni}$ mass will cause an order of magnitude increasing in the tail bolometric luminosity. The density and the size of CSM can effect the peak luminosity due to the CSM interaction. Furthermore, we compare the simulated light curves with some SNe Icn. We found that the light curves from the explosion of double WD merger remnant can fit the bolometric luminosity of SN 2019jc within $12$ days after explosion, but significantly fainter than that of SN 2019hgp and SN 2021csp. We proposed that double WD merger may be an alternative progenitor channel of at least one SN Icn. In order to better understand the progenitor systems of SN Icn and their explosion mechanisms, more SN Icn samples with high-quality data (such as early-time UV or X-ray observations) are needed in the future.

\begin{acknowledgments}

We thank the anonymous referee for his/her very helpful suggestions on the manuscript. We thank Dongdong Liu, Zhenwei Li for helpful discussion. This study is supported by the National Natural Science Foundation of China (Nos 12225304, 12288102, 12090040/12090043, 12303054), the National Key R\&D Program of China (No. 2021YFA1600404), the Yunnan Revitalization Talent Support Program -- Young Talent project, the Western Light Project of CAS (No. XBZG-ZDSYS-202117), the science research grant from the China Manned Space Project (No. CMS-CSST-2021-A12), the Frontier Scientific Research Program of Deep Space Exploration Laboratory (No. 2022-QYKYJH-ZYTS-016), and the Yunnan Fundamental Research Project (Nos 202301AU070039, 202201BC070003, 202301AU070109, 202401AU070063).
\end{acknowledgments}


\end{document}